\documentclass[conference]{IEEEtran}
\usepackage[left=54pt, right=54pt, bottom=56pt, top=54pt]{geometry} 

\IEEEoverridecommandlockouts

\usepackage{svg}
\usepackage{amsmath}
\usepackage{amssymb}
\usepackage{mathtools}
\usepackage{lipsum}
\usepackage{algorithm}
\usepackage{algorithmic}
\usepackage{xurl}
\usepackage{breakcites}
\usepackage{graphicx}
\usepackage{comment}
\usepackage{tikz}
\newcommand{\tikzxmark}{
\tikz[scale=0.23] {
    \draw[line width=0.7,line cap=round] (0,0) to [bend left=6] (1,1);
    \draw[line width=0.7,line cap=round] (0.2,0.95) to [bend right=3] (0.8,0.05);
}}
\newcommand{\tikzcmark}{
\tikz[scale=0.23] {
    \draw[line width=0.7,line cap=round] (0.25,0) to [bend left=10] (1,1);
    \draw[line width=0.8,line cap=round] (0,0.35) to [bend right=1] (0.23,0);
}}
\usepackage[dvipsnames]{xcolor}
\usepackage{tablefootnote}
\usepackage{booktabs} 
\usepackage{multirow}

\usepackage{cite}

\ifCLASSOPTIONcompsoc
  \usepackage[caption=false,font=normalsize,labelfont=sf,textfont=sf]{subfig}
\else
  \usepackage[caption=false,font=footnotesize]{subfig}
\fi

\begin{document}
%
\title{{\fontsize{16}{16}\selectfont \textbf{Distributed Incast Detection in Data Center Networks}}}
%
%
%

\author{Yiming~Zheng, 
        Haoran~Qi,
        Lirui~Yu,
        Zhan~Shu,
        and~Qing~Zhao
\thanks{Y. Zheng, H. Qi, L. Yu, Z. Shu, and Q. Zhao are with the Dept. of Electrical and Computer Engineering, University of Alberta, Edmonton, AB T6G 2R3, Canada {e-mail: zheng23, hq1, lirui1, zshu1, qingz@ualberta.ca}.}

}

\maketitle

\begin{abstract}
Incast traffic in data centers can lead to severe performance degradation, such as packet loss and increased latency. Effectively addressing incast requires prompt and accurate detection. Existing solutions, including MA-ECN, BurstRadar and Pulser, typically rely on fixed thresholds of switch port egress queue lengths or their gradients to identify microburst caused by incast flows. However, these queue length related methods often suffer from delayed detection and high error rates. In this study, we propose a distributed incast detection method for data center networks at the switch-level, leveraging a probabilistic hypothesis test with an optimal detection threshold. By analyzing the arrival intervals of new flows, our algorithm can immediately determine if a flow is part of an incast traffic from its initial packet. The experimental results demonstrate that our method offers significant improvements over existing approaches in both detection speed and inference accuracy.
\end{abstract}


%
\IEEEpeerreviewmaketitle

\section{Introduction}
The growing demand for online applications, including high-definition audio/video streaming and enterprise computing, has driven significant advancements in data center deployments. 
To meet these evolving demands, high-performance data center networks (DCNs) have been developed to support both the intensive computational workloads, such as large language model training \cite{HPN, DCN_LLM_2}, and the high-bandwidth demands of online streaming services \cite{DCN_streaming_1, DCN_streaming_2}. These innovations address the challenges of modern workloads, ensuring robust performance and scalability.


In addition to enhanced hardware capabilities, carefully designed optimization algorithms are essential for fully leveraging modern infrastructure. For example, the approach proposed in \cite{CDC-1} maximizes the quality of service while reducing energy consumption. Optimal CPU scheduling in data centers is addressed in \cite{CDC-2}, which introduces a distributed iterative algorithm. In remote direct memory access-enabled DCNs, advanced congestion control algorithms, such as HPCC \cite{HPCC}, have been developed to enhance DCN performance. However, these mechanisms still face challenges in accurately capturing and managing the full dynamics of DCNs, particularly in addressing the incast problem.

Incast, characterized by bursty traffic with high fan-in (many senders to one receiver), is both unpredictable and resource-intensive. Over the past decades, researchers have explored solutions to mitigate network degradation caused by incast, including increased packet latency and large-scale packet drops. Bursty traffic monitoring methods, such as BurstRadar \cite{BurstRadar} and BurstScope \cite{BurstScope}, can be used for incast detection and typically leverage software-defined networks (SDNs) to collect incast data across the entire DCN, enabling researchers to focus on macro-level network behaviors. Advanced data structures, like invertible sketches \cite{MV-sketch}, have also been used to enhance SDN transmission efficiency after bursts are detected. However, these centralized SDN-based burst monitoring methods often fail to effectively mitigate the immediate degradations caused by incast in DCNs, such as the well-known TCP incast problem \cite{incast_problem_1, incast_problem_2}.

Incast detection is the first step in solving the TCP incast problem. Some coarse stochastic approaches, such as sender-side pacing \cite{incast_pacing} and parameter tuning \cite{incast_parameter_tuning}, have introduced sophisticated techniques for choosing parameters like pacing rate and minimum retransmission timeout. Newer solutions leveraging advanced transmission protocols have facilitated more precise incast detection. For instance, MA-ECN \cite{MA-ECN} employs adaptive queue length thresholds for burst packet detection and marking. Pulse \cite{Pulser} and S-ECN \cite{S-ECN} utilize queue length changes (slope) as a threshold to accelerate marking in response to incast-induced burstiness. Similarly, SICC \cite{SICC} leverages queue length thresholds for incast detection as a part. Despite these advancements, relying on queue length thresholds for incast detection often results in inaccurate estimations and slow reactions \cite{qlen_thres_cons}. While using the change in queue length as a threshold can address slow reaction issues, it remains insufficient for accurate incast detection. The unpredictability of fan-in scenarios means it is extremely challenging to determine the exact number of incast flows arriving at a switch simultaneously. Consequently, this approach is prone to high detection errors and suboptimal performance (shown in our experiments).

This raises the question: \textit{Can we find a way to detect burstiness in DCNs caused by incast thoroughly and fundamentally from a probabilistic perspective.} Motivated by the above discussion, in this paper, we propose a sequential switch-based distributed hypothesis-testing method to detect incast traffic. By implementing this approach, switches can learn a priori flow characteristics from previously observed traffic patterns. Based on this knowledge, a sequential hypothesis test is performed for each new arriving flow immediately to determine whether it represents an incast scenario. 

Our method offers two primary advantages: (1) it enables high-accuracy per-flow incast detection, ensuring precise identification of incast traffic patterns; (2) it achieves fast detection by completing the process upon the arrival of the first packet in a new flow. This implies that the detection test operates independently of secondary response parameters, such as queue length or changes in queue length, which can introduce delays and inaccuracies in detection.

Experiments are conducted using the NS-3 simulator \cite{NS-3} to validate the accuracy and effectiveness of the proposed method, DIDIE. Our test suite covers diverse DCN traffic patterns, including typical scenarios such as real-world traffic distributions with large fan-in incast traffic. The results demonstrate that our hypothesis-testing approach significantly outperforms existing methods, particularly in terms of detection time and accuracy.

The rest of the paper is organized as follows: Section II explores the modeling of flow traffic and DCN topology. Section III formulates the criterion for our detection method. Section IV discusses the threshold optimization under a hypothesis testing framework, along with the methods for parameter acquisition and learning. Section V presents experimental designs and results under various traffic loads, comparing DIDIE's performance with existing methods. Section VI concludes our paper, drawing final insights and highlighting potential directions for future research endeavors.

\textbf{\textit{Notation:}} In this paper, we use the first subscript to denote traffic types, where `11' and `n1' are used for regular and incast traffic, respectively. The second subscript is used to denote the \(i\)-th traffic for its corresponding traffic type. Only incast traffic has the third subscript, representing the \(j\)-th flow within an incast traffic if required. For example, a parameter \(a\) is denoted as \(a_{11,i}\) for \(i\)-th regular traffic and \(a_{n1,i,j}\) for the \(j\)-th flow within the \(i\)-th incast traffic.

\section{System Modeling}
\subsection{Modeling of Flow Traffic}
Each server in DCNs has an `intention', either to make a request or to be requested. These `intentions' can generate traffics consisting of flows. The generated traffic can be: (1) regular traffic; (2) incast traffic. Typically, incast traffic arises from use cases such as distributed storage clusters in data centers \cite{ICTCP} or partition/aggregate traffic patterns \cite{Incast_character_1}.

\textbf{\textit{Definition 1:}} \textbf{Regular traffic} only consists of an one-to-one single flow from one sender to one receiver. \textbf{Incast traffic} consists of many-to-one multiple flows from different senders to one receiver. In addition, these multiple senders almost simultaneously initiate flows to the single receiver.

Significant research has been devoted to detecting incast traffic and mitigating its adverse effects on DCNs. However, to the best of our knowledge, existing state-of-the-art methodologies rely on secondary variables, such as switches' queue length, changes in queue length, or throughput, to indicate the presence of incast traffic, despite the availability of probabilistic a priori knowledge of traffic.

\textbf{\textit{Assumption 1:}} The inter-traffic arrival time interval for each traffic type of individual sender in DCNs follows a Poisson distribution as supported by prior works, such as \cite{Pulser, HPCC, S-ECN, PACC, incast_pacing}. Regular traffic has a higher inter-traffic arrival rate than incast traffic.

For regular traffic and incast traffic, \(t_{11,i}, t_{n1,i} \in \mathbb{R}^{+}\) denote the arrival time of the \(i\)-th traffic (\(i \in \mathbb{Z}^{+}\)) at a switch, respectively. It is evident that for all \( u, v \in \mathbb{Z}^{+} \) with \( u > v \), we have \( t_{11,u} > t_{11,v} \) and \( t_{n1,u} > t_{n1,v} \). Therefore, by Definition 1 and Assumption 1, we can formulate the following distributions:
\begin{equation} 
\nonumber
\Delta t_{11,i} \sim \operatorname{Exp}(\lambda_{11}), \ \Delta t_{n1,i} \sim \operatorname{Exp}(\lambda_{n1}), \lambda_{11} > \lambda_{n1},
\end{equation}
where the \textbf{inter-traffic arrival time interval} is denoted as \(\Delta t_{11,i} = t_{11,i} - t_{11,i-1},\) for regular traffic and, \(\Delta t_{n1,i} = t_{n1,i} - t_{n1,i-1},\) for incast traffic. The inter-traffic arrival rates are denoted as \(\lambda_{11}\) and \(\lambda_{n1}\) for regular traffic and incast traffic, respectively, where \(\Delta t_{11,i}, \Delta t_{n1,i}, \lambda_{11}, \lambda_{n1} \in \mathbb{R}^{+}.\)

By Definition 1, an incast traffic has multiple flows arriving at a switch almost simultaneously. Thus, the arrival time of the \(j\)-th flow in the \(i\)-th incast traffic, \(t_{n1,i,j}\), is given by:
\begin{multline}
\label{eq: t is monotonical increace}
t_{n1,i,j} \approx t_{n1,i,1}, \; j = 1, 2, \dots, N_{n1,i}, \\
\forall \, x,y \in \{1, 2, \dots, N_{n1,i}\}, x>y \Rightarrow t_{n1,i,x} > t_{n1,i,y},
\end{multline} 
where \(N_{n1,i}\in \mathbb{Z}^{+}\) denotes the number of flows in the \(i\)-th incast traffic. The arrival time of the first flow in an incast traffic \(t_{n1,i,1}\), is treated as the arrival time of this incast traffic \(t_{n1,i}\). The \textbf{intra-traffic arrival time interval} between two consecutive flows within the same incast traffic is defined as: \(\Delta t_{n1,i,j} = t_{n1,i,j}-t_{n1,i,j-1}.\) 

\begin{figure}[ht]
    \centering
    \includegraphics[width=3.4in]{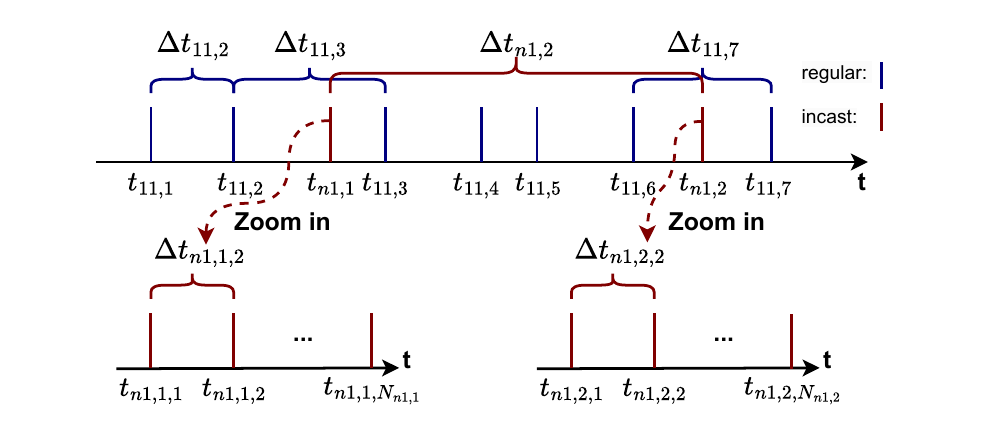}
    \caption{A Temporal Representation of Regular and Incast Traffic}
    \label{fig: timeSeries1}
\end{figure}

A schematic view of traffic arrival time at a switch port is shown in Fig. \ref{fig: timeSeries1}, where blue and red bars represent regular and incast traffic, respectively.

\subsection{Modeling of Topology}
\begin{figure}[ht]
    \centering
    \includegraphics[width=3.40in]{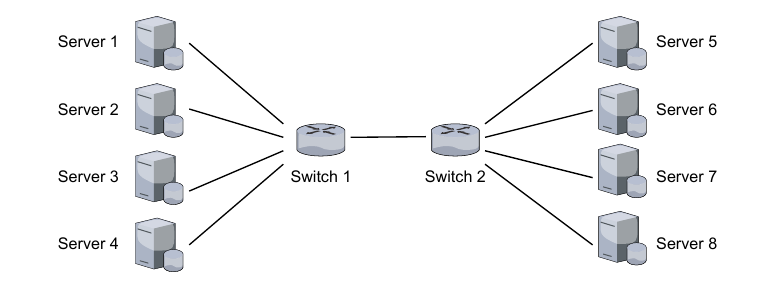}
    \caption{A 4-to-4 Dumbbell Topology}
    \label{fig:4-to-4-topo}
\end{figure}

Although our proposed method can be generalized to other topologies, we focus on the dumbbell topology, such as Fig. \ref{fig:4-to-4-topo}, for both methodology and simulation discussion due to page limitation. 
We define \(\mathcal{I}\) as the set of all possible IP addresses of servers, with cardinality \(|\mathcal{I}|\). 

\textbf{\textit{Assumption 2:}} For one traffic regardless of its type, the probability of a specific server to be the receiver with destination IP (DIP) is uniformly distributed to all IP addresses excluding the source IP (SIP).

By Assumption 2, for the \(i\)-th regular traffic, the probability of IP address of one server to be the SIP and DIP is:
\begin{multline}
\nonumber
P(\text{SIP}_{11,i} = s) = \frac{1}{|\mathcal{I}|}, \ P(\text{DIP}_{11,i} = d) = \frac{1}{|\mathcal{I}|-1}, \\ \forall s, d \in \mathcal{I}, s \neq d.
\end{multline} 

By Definition 1 and Assumption 2, for the \(i\)-th incast traffic, the probability of IP address of one server to be one of the SIPs is the following:

\begin{equation} 
\nonumber
P(\text{SIP}_{n1,i,j} = s) = \frac{N_{n1,i}}{|\mathcal{I}|}, \quad \forall s \in \mathcal{I}.
\end{equation}
All flows within an incast traffic share the same DIP, \(\text{DIP}_{n1,i} = \text{DIP}_{n1,i,j} = \text{DIP}_{n1,i,k}, \ \forall j,\, k = 1, 2, \dots, N_{n1,i}\). Let \(S \subseteq \mathcal{I} \) be the set of SIPs in the \(i\)-th incast traffic, \(S = \{\text{SIP}_{n1,i,1},\text{SIP}_{n1,i,2},...,\text{SIP}_{n1,i,N_{n1,i}} \} \). These addresses are unavailable to use as the DIP. Therefore, for incast traffic, the probability of an IP address to be DIP is: 
\begin{equation} 
\nonumber
P(\text{DIP}_{n1,i} = d) = \frac{1}{|\mathcal{I} \setminus S|} = \frac{1}{|\mathcal{I}|-N_{n1,i}}, \quad \forall d \in \mathcal{I} \setminus S.
\end{equation}

\section{Problem Formulation}
In this section, we formulate a switch-level sequential incast detection method. Each switch performs the detection upon receiving a new flow. Based on the flow arrival time, the switch can identify whether the traffic is regular or incast as soon as its first packet is received.



Flows in an incast traffic arrive at switches nearly simultaneously due to factors such as hardware variations and timing offsets. To accommodate this, we introduce a flow arrival time interval threshold \(\epsilon \in \mathbb{R}^{+}\). 

\textbf{\textit{Criterion 1: }}Let the arrival time interval between two consecutive flows be defined as \(\Delta t_k = t_k - t_{k-1}, \Delta t_k \in \mathbb{R}^{+},\) where \(t_k \in \mathbb{R}^{+}\) and \(t_{k-1} \in \mathbb{R}^{+}\) denote the arrival times of the \(k\)-th and \((k-1)\)-th flows, respectively. The index \(k \in \mathbb{Z}^{+}\) denotes time-sequenced flows received over time, regardless of traffic type. The threshold-based decision rule is:
\begin{multline}
\nonumber
\text{Classify as}
\begin{cases}
\text{incast}, & \text{if } \Delta t_k \leq \epsilon \text{ and }  \text{DIP}_k = \text{DIP}_{k-1} \\
\text{regular}, &\text{otherwise.}
\end{cases}
\end{multline}

Consider a simulation with a 4-to-1 incast traffic initiated at 1000 ns, where four distinct senders (server 1-4) concurrently transmit data flows to a single receiver (server 5) in the topology as shown in Fig. \ref{fig:4-to-4-topo}. The detection results at switch 1 for these incast flows are summarized in Table \ref{table:Performance Comparison: With vs. Without epsilon}, where servers are shortened as `S'. A check mark indicates successful detection of an incast flow, while a cross means the current flow is non-incast (regular traffic). By incorporating the threshold \(\epsilon=20\) ns, this classification method can accurately detect all incast flows even when the intra-traffic arrival times of flows in an incast traffic exhibit slight temporal offsets. 
\begin{table}[!ht]
\caption{Detection Performance with \(\epsilon=20 \ ns\)}
\label{table:Performance Comparison: With vs. Without epsilon}
\centering
\begin{tabular} { c  c  c  c  c  c  c } 
  \textbf{flow} & \boldmath{\(k\)} & \boldmath{\(t_{k}\)} & \boldmath{\(\Delta t_{k}\)} & \textbf{DIP} & \textbf{SIP} & \textbf{Result}  \\  
 \hline
 \(11,1\) & 1 & \(940\) ns & \(940\) ns  & S7 & S1 & \color{NavyBlue} \tikzxmark \\  \hline
 \(11,2\) & 2 & \(950\) ns & \(10\) ns  & S6 & S4 & \color{NavyBlue} \tikzxmark  \\  \hline
 \(n1,1,1\) & 3 & \(1000\) ns & \(50\) ns  & S5 & S1 & \! \ \({\color{NavyBlue}\tikzcmark}\)\textsuperscript{1}  \\  \hline
 \(n1,1,2\) & 4 & \(1000\) ns & \(0\) ns   & S5 & S2 & \color{NavyBlue} \tikzcmark  \\  \hline
 \(n1,1,3\) & 5 & \(1012\) ns & \(12\) ns  & S5 & S3 & \color{NavyBlue} \tikzcmark  \\  \hline
 \(n1,1,4\) & 6 & \(1015\) ns & \(3\) ns   & S5 & S4 & \color{NavyBlue} \tikzcmark  \\  
 \hline
\end{tabular} \vspace{1mm}

{ }\textsuperscript{1} The classification of this flow is corrected after its next flow is detected as incast traffic via Bayesian inference.
\end{table}

In Table \ref{table:Performance Comparison: With vs. Without epsilon}, when the first packet of flow \({k=3}\) (\(n1,1,1\)) arrives at switch 1, neither its arrival time interval nor DIP (relative to the last flow) indicates that it is part of an incast traffic. However, after receiving and classifying the subsequent flow \({k=4}\) (\(n1,1,2\)) as incast, switch 1 retroactively infers that flow \({k=3}\) is the leading flow of this incast traffic. To correctly recognize the leading flow as an incast flow, we retroactively update its classification using a backward Bayesian inference step:
\begin{equation}
\nonumber
    P( \ \text{flow}_{k-1} \in \text{incast} \ | \ \text{flow}_{k} 	\in \text{incast} \ ) = 1.
\end{equation}
In other words, once flow \({k=4}\) is identified as incast, we set flow \({k=3}\) is incast, reclassifying the leading flow correctly.

However, choosing \(\epsilon\) arbitrarily may increase the false positive (FP) rate (FPR) under Criterion 1. Consider an extreme case as shown in the Table \ref{table:A False Positive Scenario}, where flow \({k=2}\) (\(11,2\)) and flow \({k=6}\) (\(11,3\)) are incorrectly classified as part of the 3-to-1 incast traffic. This FP misclassification, indicated by red check marks, occurs because the arrival time intervals of these regular traffic flows are shorter than the threshold \(\epsilon = 20\) ns. Thus, it is necessary to optimize \(\epsilon\) for minimizing the FPR of our detection method.
\begin{table}[!ht]
\caption{Two False Positive Scenarios}
\label{table:A False Positive Scenario}
\centering
\begin{tabular} { c  c  c  c  c  c  c } 
 \textbf{flow} & \boldmath{\(k\)} & \boldmath{\(t_{k}\)} & \boldmath{\(\Delta t_{k}\)} & \textbf{DIP} & \textbf{SIP} & \textbf{Result}  \\  
 \hline
 \(11,1\) & 1 & \(950\) ns & \(950\) ns  & S7 & S3 & \color{NavyBlue} \tikzxmark  \\  \hline
 \(11,2\) & 2 & \(980\) ns & \(30\) ns  & S5 & S3 & \ \color{BrickRed} \tikzcmark \textsuperscript{1}\\  \hline 
 \(n1,1,1\) & 3 & \(1000\) ns & \(20\) ns   & S5 & S1 & \color{NavyBlue} \tikzcmark \\  \hline
 \(n1,1,2\) & 4 & \(1012\) ns & \(12\) ns  & S5 & S2 & \color{NavyBlue} \tikzcmark  \\  \hline
 \(n1,1,3\) & 5 & \(1015\) ns & \(3\) ns   & S5 & S3 & \color{NavyBlue} \tikzcmark \\  \hline
 \(11,3\) & 6 & \(1030\) ns & \(15\) ns   & S5 & S4 & \color{BrickRed} \tikzcmark \\  
 \hline
\end{tabular} \vspace{1mm}

{ }\textsuperscript{1} Incorrect Bayesian inference.
\end{table}

\section{Proposed Method}
In this section, we derive the optimal \(\epsilon\) for Criterion 1 by optimizing the receiver operating characteristic (ROC) curve based on its corresponding cost function.

\subsection{Hypothesis Test Design}
We design a hypothesis test for the latest incoming flow that has the same DIP with the second-to-last flow, where:
\begin{itemize}
    \item \textbf{Null Hypothesis} ($H_0$): The current incoming flow is regular traffic.
    \item \textbf{Alternative Hypothesis} ($H_1$): The current incoming flow is part of incast traffic.
\end{itemize}  

Under \( H_0 \) and by Assumption 1, \( \Delta t_{k} \) between consecutive flows follows an exponential distribution with rate \( \lambda_{11} \). As highlighted in Section II, the probability that two consecutive regular traffic flows have the same DIP is:
\begin{equation}
\nonumber
P^{(H_0)}\left( \text{DIP}_{\text{k}} = \text{DIP}_{\text{k-1}} \right) = \frac{1}{|\mathcal{I}|}.
\end{equation}
Therefore, the rate for two consecutive regular traffics with same DIP is \(\lambda_{11}/|\mathcal{I}|\) and the corresponding probability density function (PDF) under hypothesis \(H_0\) is given by:
\begin{equation}
\nonumber
f^{(H_0)}(\Delta t_{k}) = \frac{\lambda_{11}}{|\mathcal{I}|} e^{-\frac{\lambda_{11}}{|\mathcal{I}|} \Delta t_{k}}.
\end{equation}

Under \( H_1 \), the intra-traffic arrival time interval between two consecutive flows within an incast, \(\Delta t_{n1,i,j}\), is more tricky. The actual receiving times of these incast flows at switches are differed slightly due to hardware offsets and variants. However, to authors' best knowledge, no prior research had been conducted to address this issue. 

\textbf{\textit{Assumption 3:}} The incast receiving time offset, added to the ideal receiving time \( t_{n1,i}^* \in \mathbb{R}^{+}\), follows a half-normal distribution \( \nu_{n1,i} \sim \mathcal{N^+}(0,\sigma_{n1,i}^{2}), \sigma_{n1,i} \in \mathbb{R}^{+}.\) Furthermore, this offset deviation is much less than the expected inter-traffic arrival time of regular traffic (\(\sigma_{n1,i} \ll 1/\lambda_{11} \)). A strictly positive half-normal distribution is adopted because no actual receiving time can precede the ideal receiving time.


The set \(T_{n1,i}\) contains \(N_{n1,i}\) numbers of the observed arrival time to each flow in \(i\)-th incast traffic: \(T_{n1,i}=\{t_{n1,i,1} ,..., t_{n1,i,N_{n1,i}}\}.\) Each element in \(T_{n1,i}\) consists of two parts as defined by Assumption 3. We estimate \(t_{n1,i}^*\) and \(\sigma_{n1,i}\) based on \(T_{n1,i}\). For simplicity and computational efficiency, we estimate \(t_{n1,i}^* \approx t_{n1,i,1}.\) We use an unbiased estimator of the population variance \cite{unbias-estimator-book} to estimate \(\sigma_{n1,i}^{2}\):
\begin{equation}
\nonumber
\tilde{\sigma}_{n1,i}^{2}=\frac{1}{n-1}\sum^{N_{n1,i}}_{j=1}((t_{n1,i,j}-t_{n1,i,1})^2).
\end{equation}
So the PDF of two same DIP consecutive flows under \(H_1\) is:
\begin{multline}
\nonumber
f^{(H_1)}(\Delta t_{k}) = f^{(H_1)}(t_{n1,i,j}-t_{n1,i,1}) \\ = \frac{2}{\Tilde{\sigma}_{n1,i} \sqrt{2\pi}} e^{-\frac{(t_{n1,i,j}-t_{n1,i,1})^2}{2\Tilde{\sigma}_{n1,i}^2}}. 
\end{multline}
The likelihood ratio \( \Lambda \) and its hypothesis test are defined as:
\begin{equation}
\label{eq:final likelihood ratio}
\Lambda (\epsilon) = \dfrac{f^{(H_1)}(\epsilon)}{f^{(H_0)}(\epsilon) } = \dfrac{\frac{2}{\Tilde{\sigma}_{n1,i} \sqrt{2\pi}} e^{-\frac{\epsilon^2}{2\Tilde{\sigma}_{n1,i}^2}} }{\frac{\lambda_{11}}{|\mathcal{I}|} e^{-\frac{\lambda_{11}}{|\mathcal{I}|} \epsilon}},
\end{equation}
if \(\Lambda(\epsilon) \geq \eta,\) accept \( H_1,\) if \( \Lambda(\epsilon) < \eta, \) accept \( H_0.\)

\textbf{\textit{Lemma 1:}} Under Assumption 3, the likelihood ratio \( \Lambda \) as defined in eq. (\ref{eq:final likelihood ratio}) is monotonically decreasing, where: 
\begin{equation}
\nonumber
\frac{d\Lambda}{d\epsilon}<0, \, \forall \epsilon \geq 0.
\end{equation}

\textit{\textbf{Proof:}} The derivative of eq. (\ref{eq:final likelihood ratio}) is shown as:
\begin{equation}
\label{eq: derivative of likelihood ratio}
\frac{d\Lambda}{d\epsilon} = \Lambda \cdot (\frac{\lambda_{11}}{|\mathcal{I}|} - \frac{\epsilon}{\Tilde{\sigma}_{n1,i}^2}).
\end{equation}
It is evident that \(\Lambda(\epsilon)>0, \forall \epsilon \geq 0\). So the sign of eq. (\ref{eq: derivative of likelihood ratio}) depends on the term \((\lambda_{11}/|\mathcal{I}| - \epsilon/\Tilde{\sigma}_{n1,i}^2).\) Under Assumption 3, we have \(\lambda_{11}\Tilde{\sigma}_{n1,i}^2 / |\mathcal{I}| \approx 0.\) Thus, for any \(\epsilon > 0\), we have \(d\Lambda/d\epsilon<0\). This completes the proof of Lemma 1.

\textbf{\textit{Definition 2:}} True positive rate (TPR), denoted by \(R_{TP}\), is defined as the number of flows belonging to incast traffic that are correctly classified as incast, divided by the total number of incast flows. FPR, denoted by \(R_{FP}\), is defined as the number of flows belonging to regular traffic that are incorrectly classified as incast, divided by the total number of regular traffic flows. False negative rate (FNR) is defined as: \(R_{FN} = 1- R_{TP}\).

\textbf{\textit{Theorem 1:}} Given the eq. (\ref{eq:final likelihood ratio}), the ROC curve has the following analytical form: 
\begin{equation}
\label{eq: TPR vs FPR analytical expression}
R_{TP} = 1-2Q(-\frac{|\mathcal{I}|}{\lambda_{11} \Tilde{\sigma}_{n1,i}} \ln{(1-R_{FP})} ),
\end{equation}
where \(Q(\cdot)\) is the half-normal Q-function, defined as:
\begin{equation}
\nonumber
Q(\cdot) = \frac{2}{\Tilde{\sigma}_{n1,i} \sqrt{2\pi}} \int_{\cdot}^{\infty} e^{-\frac{\epsilon^2}{2\Tilde{\sigma}_{n1,i}^2}} \,d\epsilon.
\end{equation}

\textbf{\textit{Proof:}} By taking logs to eq. (\ref{eq:final likelihood ratio}), it gives:
\begin{equation}
\nonumber
\ln{(\Lambda)} = C - \frac{\epsilon^2}{2\Tilde{\sigma}_{n1,i}^2} + \frac{\lambda_{11}}{|\mathcal{I}|} \epsilon, \ C = \ln{(\frac{2|\mathcal{I}|}{\Tilde{\sigma}_{n1,i}\sqrt{2\pi}\lambda_{11} }) }.
\end{equation}
By Definition 2 and Lemma 1, FPR is given by:
\begin{equation}
\label{eq: FPR analytical expression}
R_{FP} = P(\epsilon \leq \gamma \, | \, H_0) = \int_{0}^{\gamma} f^{H_0}(\epsilon) \,d\epsilon = 1-e^{-\frac{\lambda_{11}}{|\mathcal{I}|} \gamma }.
\end{equation}
TPR is given by:
\begin{equation}
\label{eq: TPR analytical expression}
R_{TP} = P(\epsilon \leq \gamma \, | \, H_1) = \int_{0}^{\gamma} f^{H_1}(\epsilon) \,d\epsilon = 1-2Q(\frac{\gamma}{\Tilde{\sigma}_{n1,i}}).
\end{equation}
Thus, we can obtain \(\gamma = -\frac{|\mathcal{I}|}{\lambda_{11}}\ln{(1-R_{FP})} \) and substitute it into eq. (\ref{eq: TPR analytical expression}). This completes the proof for Theorem 1.

\begin{figure}[ht]
    \centering
    \includegraphics[width=3.4in]{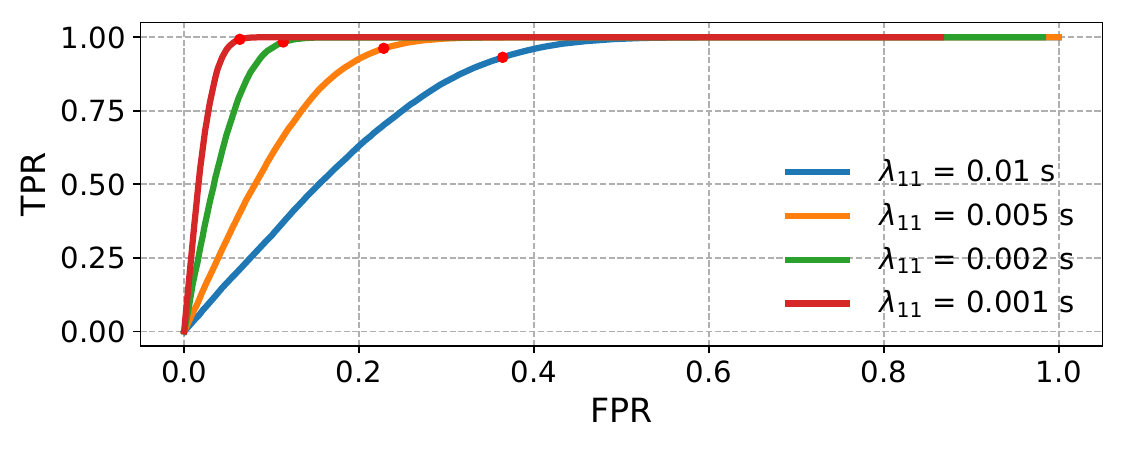}
    \caption{ROC curves for various values of \(\lambda_{11}\).}

    \label{fig:ROC curves}
\end{figure}

Based on eq. (\ref{eq: TPR vs FPR analytical expression}), ROC curves for various \(\lambda_{11}\) are shown in Fig. \ref{fig:ROC curves}. As the threshold \(\epsilon\) increases, both the TPR and FPR increase.

\subsection{Optimization of \(\epsilon\)}

After obtaining the analytical expression between TPR and FPR, the optimal threshold \(\epsilon\) can be determined based on a selected cost function. Intuitively, the optimal \(\epsilon\) corresponds to a point on the ROC curve that lies close to the top-left corner, indicating high TPR and low FPR. Various cost function formulations can be used to achieve this. In this paper, we adopt a linear cost function below:
\begin{equation}
\label{eq:linear optimization for ROC}
J(\gamma) = C_{FN} R_{FN}(\gamma) + C_{FP} R_{FP}(\gamma),
\end{equation}
where FNR and FPR costs are defined as \(C_{FN} \in \mathbb{R}^{+}\) and \(C_{FP} \in \mathbb{R}^{+}\), respectively. Then the optimal threshold \( \epsilon^* \) is given by:
\begin{equation}
\label{eq:linear optimization for ROC in short}
\epsilon^* = \arg\min_{\gamma} J(\gamma).
\end{equation}

\textbf{\textit{Theorem 2:}} Given the eq. (\ref{eq:linear optimization for ROC in short}), the optimal threshold \( \epsilon^* \in \mathbb{R}^{+}\) has a unique solution (i.e., a single stationary point), subject to the cost function in eq. (\ref{eq:linear optimization for ROC}). 

\textit{\textbf{Proof:}} By combining eq. (\ref{eq: FPR analytical expression}), (\ref{eq: TPR analytical expression}) and (\ref{eq:linear optimization for ROC}), we can obtain:
\begin{equation}
J(\gamma) = 2 C_{FN} Q(\frac{\gamma}{\Tilde{\sigma}_{n1,i}}) + C_{FP}(1- e^{-\frac{\lambda_{11}}{|\mathcal{I}|} \gamma }).
\end{equation}
The stationary points are located at: 
\begin{equation}
\label{eq: stationary point}
\frac{dJ}{d\gamma} = -\frac{2C_{FN}}{\Tilde{\sigma}_{n1,i}\sqrt{2\pi}} e^{-\frac{\gamma^2}{2\Tilde{\sigma}_{n1,i}^2}} + \frac{\lambda_{11}}{|\mathcal{I}|} C_{FP} e^{-\frac{\lambda_{11}}{|\mathcal{I}|} \gamma } = 0.
\end{equation}
By denoting \(A = \frac{2C_{FN}}{\Tilde{\sigma}_{n1,i}\sqrt{2\pi}}\), \(B = \frac{\lambda_{11}}{|\mathcal{I}|} C_{FP}\) and \(\beta = \frac{\lambda_{11}}{|\mathcal{I}|}\), eq. (\ref{eq: stationary point}), after taking logs, can be rearranged as a quadratic formula: 
\begin{equation}
\label{eq: stationary point rearraged}
\frac{1}{2\Tilde{\sigma}_{n1,i}^2}\gamma^2 - \beta \gamma + \ln{(\frac{B}{A})} =0.
\end{equation}
Therefore, the roots of eq. (\ref{eq: stationary point rearraged}) are given by:
\begin{equation}
\gamma = \Tilde{\sigma}_{n1,i}^2 (\beta \pm (\beta^2 - \frac{2}{\Tilde{\sigma}_{n1,i}^2} \ln{(\frac{B}{A})} )^{\frac{1}{2}} ), \gamma \in \mathbb{R}^{+}.
\end{equation}
Thus, we only need to show that \((\beta^2 - \frac{2}{\Tilde{\sigma}_{n1,i}^2} \ln{(\frac{B}{A})}) \geq \beta^2 \) to complete the proof. This can be reduced to show:
\begin{equation}
\ln{(\frac{B}{A})} \leq 0 \Rightarrow B \leq A \Rightarrow  \frac{\lambda_{11}}{|\mathcal{I}|} C_{FP} \leq \frac{2C_{FN}}{\Tilde{\sigma}_{n1,i}\sqrt{2\pi}}.
\end{equation}
Under Assumption 3, \( \Tilde{\sigma}_{n1,i} \ll 1/\lambda_{11} \), eq. (\ref{eq: stationary point rearraged}) has one valid real-positive solution and one invalid real-negative solution.

\begin{figure}[ht]
    \centering
    \includegraphics[width=3.4in]{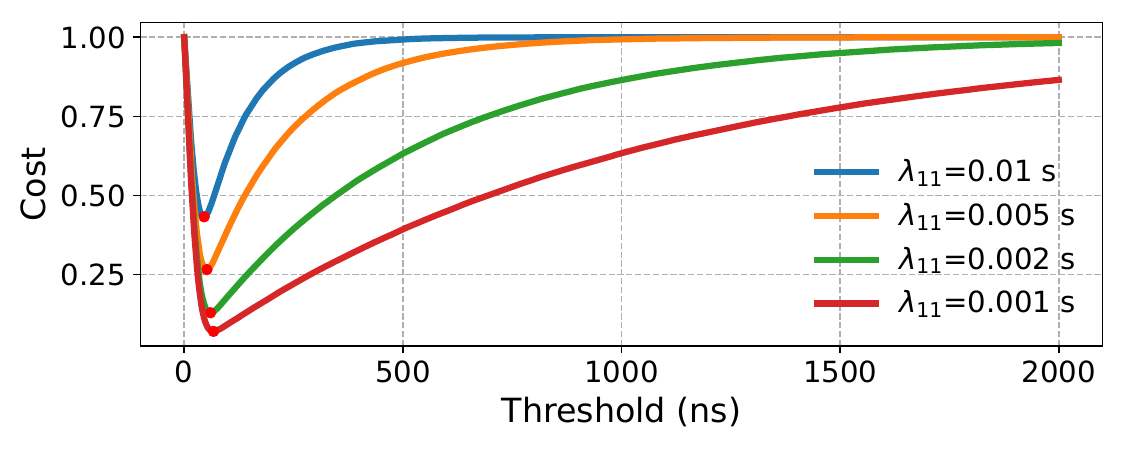}
   \caption{Cost vs. threshold for various values of \(\lambda_{11}\).}

    \label{fig: cost vs. threshold curves}
\end{figure}

As an example, let \(C_{FN}\), \(C_{FP}\), \(|\mathcal{I}|\), \(\sigma_{n1,i}\), be 10, \(10^5\), 1 and 25 ns, respectively, the resulting cost-to-threshold curves for various \(\lambda_{11}\) are plotted in Fig. \ref{fig: cost vs. threshold curves}, where the linear cost is minimized at the unique optimal \(\epsilon^*\), indicated by red points. In Fig. \ref{fig:ROC curves}, these optimal thresholds correspond to red points on the ROC curve that lie near the top-left corner, indicating a favorable balance of high TPR and low FPR. The computed optimal thresholds \(\epsilon^*\) are: 45.6 ns, 51.9 ns, 60.1 ns, and 67.0 ns for \(\lambda_{11}=\) 0.01 s, 0.005 s, 0.002 s, and 0.001 s receptively.

\textit{\textbf{Remark 1: }}Note that eq. (\ref{eq:final likelihood ratio}) is used to identify the first two leading flows of an incast traffic, where their previous flow is regular traffic. If the detection applies to subsequent intra-traffic flows in an incast traffic, eq. (\ref{eq:final likelihood ratio}) shall be modified as:
\begin{equation}
\label{eq:final likelihood ratio in the middle of incast}
\Lambda (\epsilon) = \dfrac{f^{(H_1)}(\epsilon)}{f^{(H_0)}(\epsilon) } = \dfrac{\frac{2}{\Tilde{\sigma}_{n1,i} \sqrt{2\pi}} e^{-\frac{(\epsilon+t_{n1,p,m}-t_{n1,p,1})^2}{2\Tilde{\sigma}_{n1,i}^2}} }{\frac{\lambda_{11}}{|\mathcal{I}|} e^{-\frac{\lambda_{11}}{|\mathcal{I}|} (\epsilon+t_{n1,p,m}-t_{11,q})}},
\end{equation}
where \(p\) and \(q\) denote the latest indices for incast and regular traffic, respectively, and \(m\) represents the latest flow received within this \(p\)-th incast traffic so far. Thus, \((\epsilon+t_{n1,p,m}-t_{n1,p,1})\) is used instead of \(\epsilon\) to represent the interval since the leading flow of the \(p\)-th incast traffic under \(H_1\). Similarly, \((\epsilon+t_{n1,p,m}-t_{11,q})\) is used instead of \(\epsilon\) to represent the time interval since the latest \(q\)-th regular traffic under \(H_0\). However, for a \(p\)-th incast traffic that occurs between two regular traffics, given that the expected time interval between two consecutive regular flows with the same DIP satisfies, \(|\mathcal{I}|\lambda_{11} > \mathbb{E}[t_{n1,p,m}-t_{11,q}] \gg \mathbb{E}[t_{n1,p,m}-t_{n1,p,1}],\) the term \(|\mathcal{I}|/\lambda_{11}\) dominates \(\Lambda (\epsilon)\) more than two exponential terms (the difference \(t_{n1,p,m}-t_{n1,p,1}\) caused by the hardware offset is a small number compared to \(|\mathcal{I}|/\lambda_{11}\)). 

Thus, for all subsequent discussions, we adopt eq. (\ref{eq:final likelihood ratio}) for computing \(\epsilon^*\), rather than eq. (\ref{eq:final likelihood ratio in the middle of incast}), even for non-two leading flows in an incast traffic. This approximation eliminates the requirement for online computation of \(\epsilon^*.\)

\subsection{Parameter Acquisition}
Next, we describe the procedure for estimating the parameters required for the proposed hypothesis testing framework.

The cardinality \(|\mathcal{I}|\) is determined by the number of servers, which is known from the DCN topology and structure. The arrival time interval \(\Delta t_{k}\) is computed by taking the difference between the arrival times of the leading packets of the two most recent flows. The DIPs are extracted from the packet headers. The remaining parameters to be determined are the inter-traffic arrival rate for regular traffics \(\lambda_{11}\) and the variance of the half normal disturbance of incast traffics \(\sigma_{n1,i}^{2}\). 

Our method `learns' \(\lambda_{11}\) sequentially based on the past observations of \(\Delta t_{11,i}\). By defining the most recent rate observation as \( \lambda_{11,i} = 1/ \Delta t_{11,i}\), we employ an exponentially weighted moving average (EWMA) method to sequentially measure \(\Tilde{\lambda}_{11,i}\) for estimating \(\lambda_{11}\) with weight \(\alpha, \ \forall a \in (0,1)\):
\begin{equation}
\lambda_{11} \approx \Tilde{\lambda}_{11,i} = \alpha  \lambda_{11,i} + (1-\alpha) \Tilde{\lambda}_{11,i-1}.
\end{equation}

Since the hardware offset that differs the arrival times of intra-traffic flows are expected to be the same across all incast traffics, we approximate \(\sigma_{n1,i}^2 \approx \sigma_{n1}^{2} \). The similar EWMA method is used, where:
\begin{equation}
\sigma_{n1,i}^2 \approx \sigma_{n1}^{2} = \alpha \Tilde\sigma_{n1,i}^{2}  + (1-\alpha) \sigma_{n1,i-1}^{2}.
\end{equation}

The above estimations, \(\Tilde{\lambda}_{11,i}\) and \(\sigma_{n1}^{2}\), converge to their ground truth values, \(\lambda_{11}\) and \(\sigma_{n1,i}^2\), respectively, as more flows are received by the switches. Additionally, since \(\sigma_{n1,i}^2\) remains approximately constant, its estimation can also be performed offline to reduce computational cost.

\section{Experiments}
In this section, our proposed method, DIDIE, is validated through experiments and compared against the queue length and queue length gradient threshold methods as two baselines. The experiments are conducted on the dumbbell topology as shown in Fig. \ref{fig:4-to-4-topo}, with 1 Gbps link rate and 1,000 ns link delay. The packet size is set to 1,100 bytes.

\subsection{Incast Traffic-Only Experiment}
In this experiment, we inject incast traffics with various sizes, ranging from 2-to-1 to 4-to-1, to evaluate the performance of DIDIE and the two baselines. Since this is an incast-only experiment, we set \(\lambda_{11} = 0\). The resulting optimal threshold for DIDIE is determined as \(\epsilon^* \approx 14\) ns. 

The results are summarized in Table \ref{tab:Performance comparison of methods across thresholds}, demonstrating that DIDIE accurately captures incast traffics of different sizes with minimal detection time at switches. In contrast, the queue length threshold method (with a threshold of 10 KB) requires significantly more detection time, as the queue length takes time to reach its threshold. The queue length gradient threshold method (with a threshold of 2 GB/s) detects incast traffic faster than the queue length threshold method. However, it fails to detect 2-to-1 incast traffics, as the queue length increase rate under this scenario is below the 2 GB/s threshold.

\begin{table}[ht]
\centering
\caption{Performance comparison of methods across thresholds.}
\label{tab:Performance comparison of methods across thresholds}
\begin{tabular}{ccccc}
\textbf{Method} & \textbf{incast}  & \textbf{Threshold}  & \textbf{Detected ?} & \textbf{ Time}  \\
\midrule
\multirow{2}{1.3cm}  {DIDIE}       & 2-to-1 & 14 ns   &\color{NavyBlue}\tikzcmark   & \textbf{1.2 ns}   \\
                                   & 3-to-1 & 14 ns  & \color{NavyBlue}\tikzcmark   & \textbf{1.2 ns}  \\
                                   & 4-to-1 & 14 ns  & \color{NavyBlue}\tikzcmark   & \textbf{1.2 ns}  \\
\midrule
\multirow{3}{1.3cm}  {QLen Threshold} & 2-to-1 & 10 KB   & \color{NavyBlue}\tikzcmark & 88,000 ns   \\
                                      & 3-to-1 & 10 KB  & \color{NavyBlue}\tikzcmark  & 44,000 ns  \\
                                      & 4-to-1 & 10 KB  & \color{NavyBlue}\tikzcmark  & 25,200 ns  \\
\midrule
\multirow{3}{1.3cm}  {\(\dot{\text{QLen}}\) Threshold} & 2-to-1 & 2 GB/s   & \color{BrickRed}\tikzxmark &                                                            N/A   \\
                                                       & 3-to-1 & 2 GB/s  & \color{NavyBlue}\tikzcmark  & 17,600 ns  \\
                                                       & 4-to-1 & 2 GB/s  & \color{NavyBlue}\tikzcmark  & 17,600 ns  \\
\end{tabular}
\end{table}

\subsection{Incast Traffic with Real-World Traffic Experiment}
In this experiment, we run 4-to-1 incast traffics (\(\sigma_{n1} = 3 \) ns, 2.5\% sender's load, 47 KB per incast flow) with a real-world regular traffic distribution, \(FB\_Hadoop\) \cite{FB_distribution}, under 30\% and 60\% sender's load. The expected regular traffic arrival time interval can be calculated as, \( 1/\lambda_{11} =  \text{averageFlowLength} / \text{linkRate} / \text{load}\% \). Thus, \(\lambda_{11} = \) 3.2 ms and 1.6 ms for 30\% and 60\% sender loads, respectively. The optimal threshold for DIDIE is determined as \(\epsilon^* = 14\) ns. 

As shown in Table \ref{tab:Real-World Traffic Performance Comparison.}, DIDIE achieves the minimal average incast detection time in a real-world \textit{FB\_Hadoop} traffic scenario mixed with incast traffics, while maintaining 0\% FPR under both 60\% and 30\% traffic loads. 

For queue length threshold method, a 10 KB threshold requires more detection time than a 5 KB threshold. However, lowering the threshold increases the FPR due to the presence of the background regular traffic from \textit{FB\_Hadoop}. This also explains why the FPR increases as the \textit{FB\_Hadoop} load\% increases, even with the same queue length threshold. The detection time decreases as the \textit{FB\_Hadoop} load\% increases since the queue length is partially stacked by this background traffic. 

For queue length gradient method, the average detection time is lower than that of the queue length method. However, the FPR is higher, as the queue length gradient is more easily affected by regular traffic from \textit{FB\_Hadoop}. Additionally, the FPR further increases with either a higher \textit{FB\_Hadoop} load\% or a lower queue length gradient threshold.

\begin{table}[h]
\centering
\caption{Real-World Traffic Performance Comparison.}
\label{tab:Real-World Traffic Performance Comparison.}
\begin{tabular}{ccccc}
\textbf{Method} & \textbf{\(FB\_Hadoop\)}  & \textbf{Threshold}  & \textbf{Avg Time}  & \textbf{FPR}  \\
\midrule
\multirow{3}{1.3cm}{DIDIE}       & 60\% & 14 ns     & \textbf{1.2 ns} &  \textbf{0\%}\\
                                   & 30\% & 14 ns    & \textbf{2.7 ns} & \textbf{0\%}\\
                                   
\midrule
\multirow{3}{1.3cm}  {QLen Threshold} & 60\% & 10 KB     & 31,766 ns & 14.3\% \\
                                      &  & 5 KB     & 14,167 ns &  25.0\%\\
                                      & 30\% & 10 KB      & 32,263 ns & 0.0\%\\
                                      &  & 5 KB      & 19,062 ns & 14.3\%\\
\midrule
\multirow{3}{1.3cm}  {\(\dot{\text{QLen}}\) Threshold} & 60\% & 3 GB/s    & 14,169 ns & 25.0\% \\
                                                       &  & 1 GB/s    & 14,050 ns & 33.3\%\\
                                                       & 30\% & 3 GB/s    & 17,598 ns & 0.0\%\\
                                                       &  & 1 GB/s    & 17,598 ns & 14.3\%\\
\end{tabular}
\end{table}

\section{Conclusion}\label{ch:Conclusions and Future Work}
   In this paper, a distributed switch-level incast detection method through a sequential hypothesis test has been designed for identifying incast traffic from regular traffic. The analytical solution for the optimal flow arrival time interval threshold has been derived by solving a linear cost function with the ROC curve. By adopting this optimal value, our experiments have demonstrated that the proposed method significantly outperforms existing queue length and queue length gradient threshold methods in both detection time and accuracy, without incurring additional computational cost. Future work will focus on implementing our method into other network mechanisms utilizing incast detection, such as incast-aware pacing and congestion control.

\bibliographystyle{IEEEtran}
\bibliography{IEEEexample.bib}

\end{document}